\documentclass{article}
\pdfoutput=1

\usepackage[pdftex]{graphicx}
\usepackage{amsmath}
\usepackage{amssymb}

\newcommand{\esp}{\mathsf{E}}
\newcommand{\var}{\mathsf{var}}

\newcommand{\bd}{\boldsymbol}

\title{Posterior probability of the Likelihood Ratio and (Fractional) Bayes Factor: new theoretical relations and practical uses.}

\author{I. Smith and A. Ferrari\\
Lab. H. Fizeau, \\
UNS, CNRS,
OCA\\
Campus Valrose, F-06108 Nice cedex, France\\
\texttt{isabelle.smith@unice.fr, ferrari@unice.fr} 
 }

\begin{document}

\maketitle

\begin{abstract}
In the simple vs composite hypothesis test with a proper prior, the Bayes Factor (BF) is shown to be the posterior mean of the Likelihood Ratio (LR). Therefore, the posterior standard deviation of the LR or rather its posterior cumulative density function can be used to indicate the significativity of a detection by the BF and this detection procedure can be computed from a single Markov Chain. It is applied and compared for exoplanet detection.

The previous statistics can be expressed from the Fractional BF (FBF) \cite{ohagan95} and the Probability distribution of the LR (PLR) \cite{aitkin97}. Two properties of the PLR related to the GLRT are noted and a procedure to optimize the PLR and the FBF two-parameters detectors according to their ROC curves is proposed. The performances of all tests are compared.
\end{abstract}

\section{Introduction}
\label{sec_intro}

The detection of a signal from low signal-to-noise ratio data is a general issue in signal processing. For a given dataset $\bd x$, we express the detection as the deterministic choice among a simple (no signal: $\bd \eta_0 = \bd 0$) and a composite hypothesis test:
\begin{align}
H_0: \bd \eta = \bd \eta_0 \qquad H_1: \bd \eta \sim \pi(\bd \eta)
\label{eq_hyp_test}
\end{align}
$\pi(\bd \eta) \neq \delta(\bd \eta)$ is a given \textbf{proper} multivariate prior describing the uncertainty and constraints on the intensities $\bd\eta \in \mathbb R^L$ of the signal of interest. Alternatively to the 0-1 decision an interesting no-decision region could have been used \cite{berger94,dempster08}. 

The likelihoods have the same expression under $H_0$ and $H_1$ and depend on $\bd\eta$ only. This means for frequentists that all other parameters are known. For Bayesianists, they have been marginalized out. \\

In addition to the mere detection result, information about the significativity of the decision is in general expected. In frequentist settings it is usually given by the PFA or the p-value of the statistics of detection \cite{lehmann05}. These notions are also studied in the Bayesian perspective \cite{vlachos03,bayarri98}. However, the PFA as well as the p-value require an integration of the likelihood (or other Bayesian distributions) over a subset of the sample space and this computation may be intractable.

In Bayesian settings, the Posterior Odds Ratio $\text{POR} = \frac{\text{Pr}(H_0 | \bd x)}{\text{Pr}(H_1 | \bd x)}$ minimizes the Bayesian risk under the 0-1 loss function and appears as the expression of what is exactly looked for. It is equal to the classical Bayes Factor (BF) \cite{kass95}  
\begin{align}
\text{BF} = \frac{p(\bd x | H_0)}{p(\bd x | H_1)} = \frac{p(\bd x | \bd\eta = \bd \eta_0)}{\int d \bd\eta ~ p(\bd x | \bd\eta) p(\bd \eta)}    
\label{def_bf}
\end{align}   
up to the multiplicative prior odds ratio $\text{pOR} = \text{Pr}(H_0)\text{Pr}(H_1)^{-1}$. 

In \cite{berger94}, the Bayesian detector consists in thresholding the BF and giving as the error the posterior probability of the selected model Pr$(H_{d(\bd x)} | \bd x)$ because it gives an "intrinsic significance level" \cite{birnbaum62}. However, for $i\in\{0,1\}$
\begin{align}
\text{Pr}(H_i | \bd x) = \frac{p(\bd x | H_i)\text{Pr}(H_i)}{\sum_{j=1}^2 p(\bd x | H_j)\text{Pr}(H_j) } = \frac{1}{1 + (\text{BF}\times \text{pOR})^{1-2\delta_{0i}}}  \nonumber
\end{align}
where $\delta_{ij}$ is the Kronecker symbol. The two pieces of information delivered by the detector and the "error" are largely redondant since their relation involves no other quantity than the pOR. Consequently, we consider them as insufficient outputs of the detection procedure. \\

Another important issue is the performance that can be reached by the detector. Following naturally from the first following study, both issues will be adressed theoretically and practically. \\

\section{New practical error inference: \\for a detection from the Bayes Factor}
\label{sec_stats}

In the simple versus composite test (\ref{eq_hyp_test}), when there is no nuisance parameter or when they are marginalized out thanks to a Bayesian computation, the (Bayesian) Likelihood Ratio is:
\begin{equation}
\text{LR}(\bd \eta) = \frac{p(\bd x | \bd\eta = \bd \eta_0)}{p(\bd x | \bd\eta)}
\label{def_LR}
\end{equation}
The dependencies on $\bd x$ are dropped in the sequel. For a given $\bd x$, it is a function of only one random parameter and has therefore a posterior distribution under $H_1$. 

If the prior $\pi$ is proper, it turns out (no reference found) that the Bayes Factor (\ref{def_bf}) is equal to the posterior mean of the LR:
\begin{align}
\text{BF} &= \frac{p(\bd x | H_0)}{p(\bd x | H_1)} \int d\bd \eta ~ \pi(\bd \eta) \nonumber \\
&= \frac{p(\bd x | H_0)}{p(\bd x | H_1)} \int d\bd \eta ~ \frac{\pi^*(\bd \eta | \bd x) p(\bd x | H_1)}{p(\bd x | \bd \eta)} \nonumber \\
  & = \esp^{\pi^*}[\text{LR}(\bd\eta) | \bd x]  ~~~~~~~~~~~~~~~~\text{where } \pi^* \text{is the posterior of $\bd \eta$}
  \label{prop_bf}
\end{align}
Uncertainty on the detection could then naturally be given by the posterior standard deviation of the LR: 
\begin{align}
\text{"LR}(\bd \eta) &= \text{BF} \pm \hat\sigma"   \label{eq_uncertainty} \\
\text{with }~~ \hat\sigma = (\var^{\pi^*}[\text{LR}(\bd \eta) &| \bd x])^{1/2} = (\text{FBF}(-1) - \text{BF}^2 )^{1/2}
\label{def_err}
\end{align}
where we recognized the Fractional Bayes Factor \cite{ohagan95}
\begin{align}
\text{FBF(b)} &= \frac{p(\bd x | \bd \eta=\bd \eta_0)}{\int d\bd\eta ~ p(\bd x | \bd \eta)\pi(\bd\eta)}\left(\frac{p(\bd x | \bd \eta=\bd \eta_0)^b}{\int d\bd\eta~ p(\bd x | \bd \eta)^b \pi(\bd\eta)}\right)^{-1} \nonumber \\
  &= \esp^{\pi^*}[\text{LR}(\bd \eta)^{1-b}|\bd x] 
\label{def_FBF} 
\end{align}
except that the FBF has initially been proposed for $b\in[0,1)$ as a partial Bayes Factor developped to extend the BF to improper priors. 

However, the BF is used as a statistics to threshold: the underlying distribution of LR($\bd\eta$) is explored in a non symmetric fashion and the uncertainty (\ref{eq_uncertainty}) related to the 2nd moment may be inappropriate. 

An alternative is the computation of a confidence interval or simply of the cumulative distribution of the variable: 
\begin{align}
\text{PLR}(\zeta)=\text{Pr}^{\pi^*}\{\text{LR}(\bd \eta) \le \zeta | \bd x\}  
\label{def_PLR}
\end{align}
It turns out that the Posterior distribution of the LR has also already been slightly studied. It has been proposed in \cite{dempster74} and extended and applied in \cite{aitkin97,aitkin05}. However its use could be more advocated.\\

For a practical use of the suggested tools, we propose to use a single Monte Carlo Markov Chain $\bd \eta^{[n]}\sim\pi^*(\bd\eta|\bd x)$ for all estimation and detection purposes:
\begin{itemize}
\item the chain LR($\bd \eta^{[n]}$) is straightforwardly computed 
\item the FBF (\ref{def_FBF}) is computed from an importance sampling procedure (e.g. a simple average) from LR$(\bd \eta^{[n]})$ for a given $b$. FBF is used for BF = FBF(0) and possibly for $\hat\sigma$ (\ref{def_err}).
\item the PLR (\ref{def_PLR}) is computed as the empirical cumulative distribution of the LR chain. 
\item if a signal is detected, $\bd \eta^{[n]}$ can be finally used for estimation 
\end{itemize}

\section{Other Bayesian detectors related to the Posterior probability of the LR}

The PLR (\ref{def_PLR}) and FBF (\ref{def_FBF}) appeared in Sec. \ref{sec_stats} as natural statistics for the definition of a coherent procedure for detection and have proved easy to compute numerically. They are further studied here.

\subsection{Properties of the PLR (related to the GLRT)}
\label{sec_plr_glrt}

First, we make and show two general remarks (no reference found) about the posterior density $p_{\text{LR} | \bd X}$ of the LR: 
\begin{itemize}
\item The minimum of its support is the GLRT: $$\min_\zeta\{\zeta: p_{\text{LR} | \bd X}(\zeta | \bd x) > 0\} = \text{GLRT}$$
\item Under regularity assumptions that get stronger as $L$ (the size of $\bd \eta$) increases, the function $\zeta \rightarrow p_{\text{LR} | \bd X}(\zeta | \bd x)$ diverges for $\zeta \rightarrow \text{GLRT}^+$ . 
\end{itemize}

In the same Bayesian frame as for the definition of LR (\ref{def_LR}), the (Bayesian) Generalized Likelihood Ratio Test (GLRT) is defined for the simple versus composite test (\ref{eq_hyp_test}) by 
\begin{equation}
\text{GLRT} = \min_{\bd\eta\in\mathcal{E}} ~\text{LR}(\bd \eta) \label{def_GLRT}
\end{equation}
where $\mathcal E = \text{Sup}(p_{\bd x | \bd \eta}) \cap \text{Sup}(\pi)$ in order to take into account the definition domain of the likelihood and the constraints of the parameter set. Therefore, $\mathcal E = \text{Sup}(\pi^*)$ and
\begin{align}
\text{Pr}^{\pi^*}\{\text{LR}(\bd \eta) < \text{GLRT} | \bd x\} = 0
\end{align}

Under regularity assumptions in the neighborhood of the GLRT (reached by definition for $\bd \eta = \hat{\bd \eta}_{\text{ML}}$) we have
\begin{align}
p_{\text{LR} | \bd X}(\zeta | \bd x) \rightarrow \infty ~~\text{ when } ~~ \zeta \rightarrow \text{GLRT}^+ 
\label{eq_pdf_GLRT}
\end{align}

In the following, we drop the conditionality on $\bd x$. A usual transform to infer the distribution of LR is 
$\phi: \bd \eta \rightarrow$ (LR, $\check{\bd \eta}_1$) where we note $\check{\bd \eta}_1 = ( \eta_2,..,\eta_L)$. Its Jacobian determinant is $|J| = |\partial \text{LR} / \partial \eta_1|$. The usual variables transformation gives for an open set
\begin{equation}
p_{\text{LR},\check{\bd \eta}_1}(\zeta,\check{\bd u}_1) = \sum_{k=1}^{n(\zeta,\check{\bd u}_1)} p_{\bd \eta}(\bd u^k) \left|\frac{\partial \text{LR}}{\partial \eta_1}(\bd u^k)\right|^{-1} \nonumber
\end{equation}
where the $\bd u^k$ are the solutions of $\phi(\bd u^k)=(\zeta,\check{\bd u}_1)$.

For $L =1$ ($\bd \eta$ is scalar), it gives directly the result: if the function LR : $\eta \rightarrow \zeta$ is continuously differentiable, $|d\text{LR}/d\eta|(u) \rightarrow 0$ as $u \rightarrow \text{arg}\min(\text{LR}(u))$. So $p_{\text{LR}}(\zeta) \rightarrow \infty$ as $\zeta \rightarrow \text{GLRT}$.

For $L>1$, $L-1$ integrations are required to marginalize out $\check{\bd \eta}_1$. They have to be computed for a given $\zeta >$ GLRT since the Jacobian is not defined at $\zeta = $ GLRT and since we assume $\hat{\bd \eta}_{\text{ML}}$ is the only solution of LR($\hat{\bd \eta}_{\text{ML}}$) = GLRT so that the integrand would be positive on a null set only. We show \cite{smith09d} that if locally there exist $\alpha > L$ and $(\alpha_1,..,\alpha_L)\in\mathbb R ^{L}_{+*}$ such that for all $\bd \eta$ close enough to $\hat{\bd \eta}_{\text{ML}}$
\begin{equation}
\text{LR}(\bd \eta) \le \text{GLRT} + \sum_{\ell=1}^L\alpha_\ell (\bd \eta-\hat{\bd \eta}_{\text{ML}})_\ell^\alpha \nonumber
\end{equation}
then $\epsilon^{-1}\text{Pr}(\text{GLRT} < \text{LR} \le \text{GLRT} + \epsilon) \rightarrow \infty$ when $\epsilon \rightarrow 0$.

\subsection{Optimal parametrization of PLR and FBF}
\label{sec_optim}

In addition to their initial developments motivations, the PLR and FBF are interesting to study as detectors because they can be seen as families of tests parametrized by two parameters:
\begin{align}
&\text{Reject } H_0 \text{ if  ~~~ PLR}(\zeta_0) > p_0 \nonumber \\
&\text{Reject } H_0 \text{ if ~~~  FBF}(b) < \zeta \nonumber
\end{align}
For PLR, $\bd \lambda = (\zeta_0, p_0)$ and for FBF $\bd \lambda' = (b,\zeta)$. Unlike detectors defined from a single threshold, it is possible to optimize each family. We propose to do it using the frequentist ROC curve tool, which displays the Probability of good Detection (PD) as a function of the Probability of False Alarm (PFA). For the PLR, 
\begin{align}
\text{PFA}(\zeta_0,p_0) = \text{Pr}^{p(\bd x | H_0)}\{\text{PLR}(\zeta_0) > p_0\}
\end{align}

In principle, the idea is first to compute PFA($\bd \lambda$) and PD($\bd \lambda$) for all $\bd \lambda = (\zeta_0, p_0)$, then, fix a PFA$_0$, obtain the corresponding $\{\bd \lambda: \text{PFA}(\bd\lambda) = \text{PFA}_0\}$ curve and choose from it $\bd \lambda_{\text{opt}}(\text{PFA}_0)$ that maximizes PD($\bd\lambda$). We propose to do it numerically thanks to the practical computation proposed in Sec. \ref{sec_stats}. From a large number of datasets (one set of datasets under $H_0$ and one set of datasets under $H_1$), two matrices made of the LR chains are formed. The trick is that PFA, PD and $p_0$ are asymptotically regularly sampled in the matrices as soon as the matrices are reordered. Then, the approximate optimal parameters can almost be "read" from the tables.

\section{Application of the detection procedure}

The estimation-detection procedure of Sec. \ref{sec_stats} is realistically applied to the detection of exoplanets from direct imaging using the future VLT instrument SPHERE.

\subsection{Statistical model for exoplanet detection in direct imaging}
\label{sec_model}

To specify the statistics BF etc as pure functions of $\bd x$, a statistical model is required. A hierarchical Bayesian model precisely related to our context has been developped in \cite{smith09b} and is summed up here.

The $\bd \eta$ vector of the hypothesis test (\ref{eq_hyp_test}) refers to the exoplanet intensity in the different channels. The marginalized prior $\pi(\bd \eta)$ has a positive support, is proper and approximatively scale invariant. 

The dataset is made of $K$ successive sets of $L$ images, where each image is a $M \times 1$ vector $\bd i_\ell(k)$. The $\bd x_k^t = (\bd i_1(k)^t,.., \bd i_L(k)^t)$ are assumed to be conditionally independent and described by:
\begin{equation}
\bd x_k|\bd \mu,\Sigma,\bd \eta \sim \mathcal{N}_{LM}(A_k \bd \eta +\bd \mu,\Sigma)   \label{likelihood}
\end{equation}
$A_k$, the source profiles, are assumed to be known. This first level likelihood is marginalized using conjugate priors and leads to an explicit form for $p(\bd x | \bd \eta)$ where $\bd x = \{\bd x_k\}_{k=1,..,K}$. 

The Markov chain $\bd \eta^{[n]} \sim \pi^*(\bd \eta | \bd x)$ necessary to compute PLR and FBF (see Sec. \ref{sec_stats}) is obtained from a slice sampling method \cite{robert99}.

\subsection{Application of the detection procedure on a realistic dataset}

\begin{figure}
\centerline{
\includegraphics[width=0.3\columnwidth]{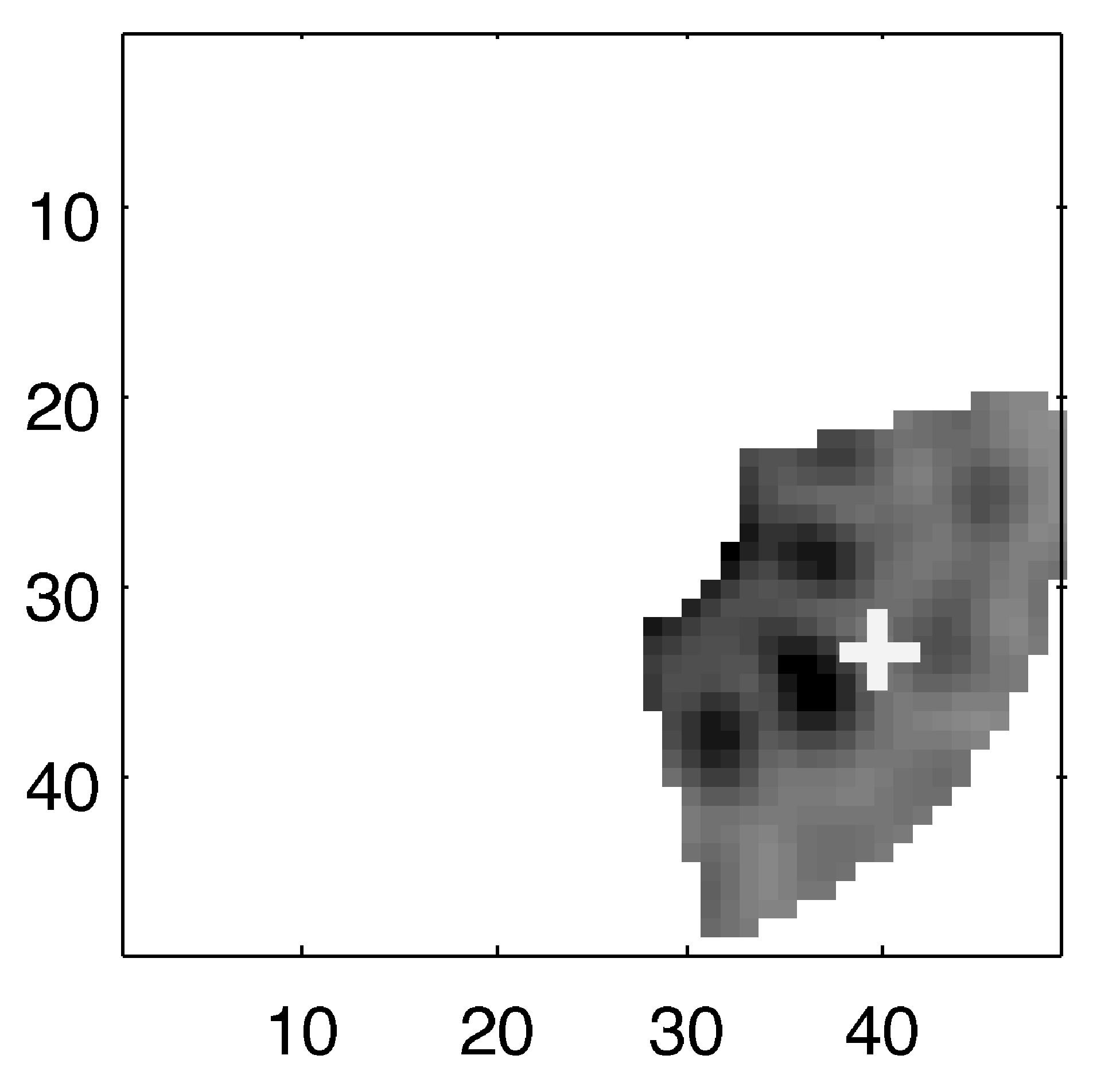}~~~
\includegraphics[width=0.31\columnwidth]{figure1.pdf}}
\vspace{-0.1cm}
\caption{Simulated data from CAOS-SPHERE with a contrast of $10^6$ between the star and the planet. Left: $\bd x_{2}(20)^{0.2}$. Right: $\bd p_{2}(20)^{0.2}$. }
\label{fig_true_data}
\end{figure}

The simulation of realistic astrophysical datasets is performed by the dedicated physical step-by-step Software Package SPHERE \cite{carbillet08} developed and used within the CAOS environment \cite{carbillet04}. A dataset $\bd x$ is simulated under $H_1$ with a luminosity contrast of $10^6$ between the star and the exoplanet (corresponding to an intensity $\bd \eta_{H1}$), and another under $H_0$, obtained from an area adjacent to the one under $H_1$. The data under $H_1$, of size $(K,L,M) = (20,2,425)$ are illustrated on Fig. \ref{fig_true_data}. Note that it is impossible to simulate many datasets.

The detection procedure described in Sec. \ref{sec_stats} and used with the realistic statistical model summarized in Sec. \ref{sec_model} is finally applied to these two datasets. The hyperparameters are chosen simply ($\nu = 2M$, $\Sigma_0 = \widehat{\sigma^2} I_{2M}$ ...) or unfavourable ($\bd m_0 = \ln(1000\bd\eta_{H1})$). Both chains $\eta^{[n]}$ are made of $N=$10$^5$ samples.

\begin{figure}
\centerline{\includegraphics[width=1\columnwidth]{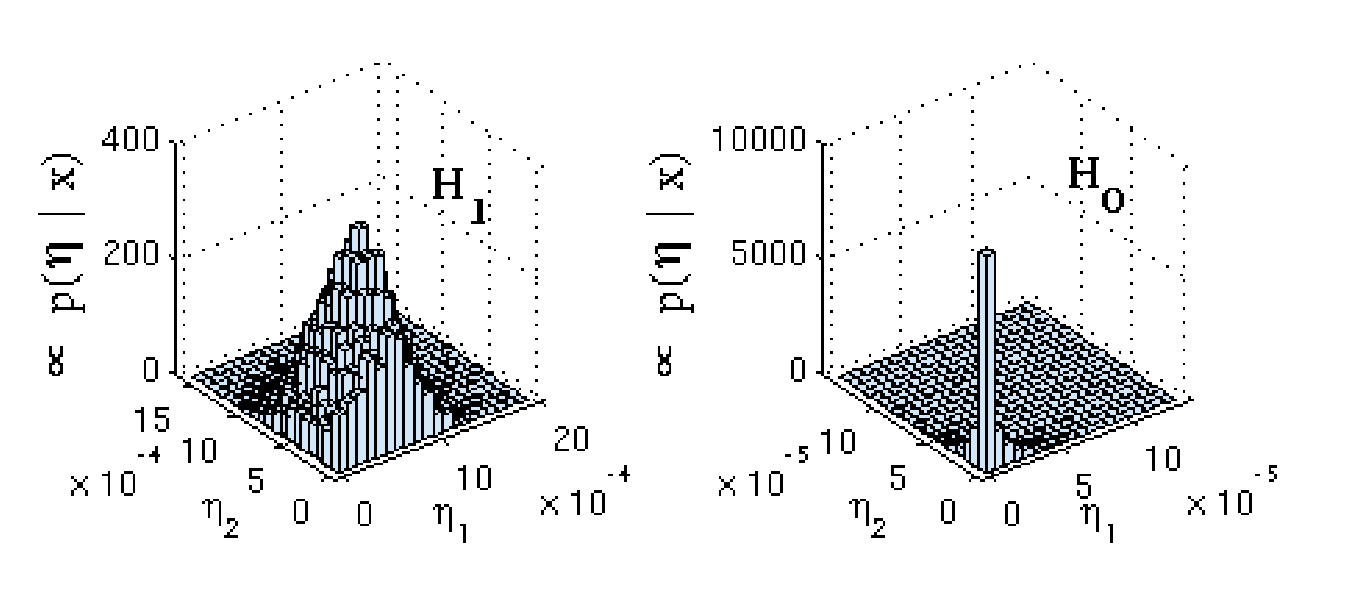}}
\vspace{-0.3cm}
\caption{Histograms of two Markov chains $\bd \eta^{[n]} \sim p(\bd \eta|\bd x)$ resulting from the data with ($H_1$, left) and without ($H_0$, right) a planet .}
\label{fig_posterior}
\end{figure} 

\begin{figure}
\centerline{\includegraphics[width=1\columnwidth]{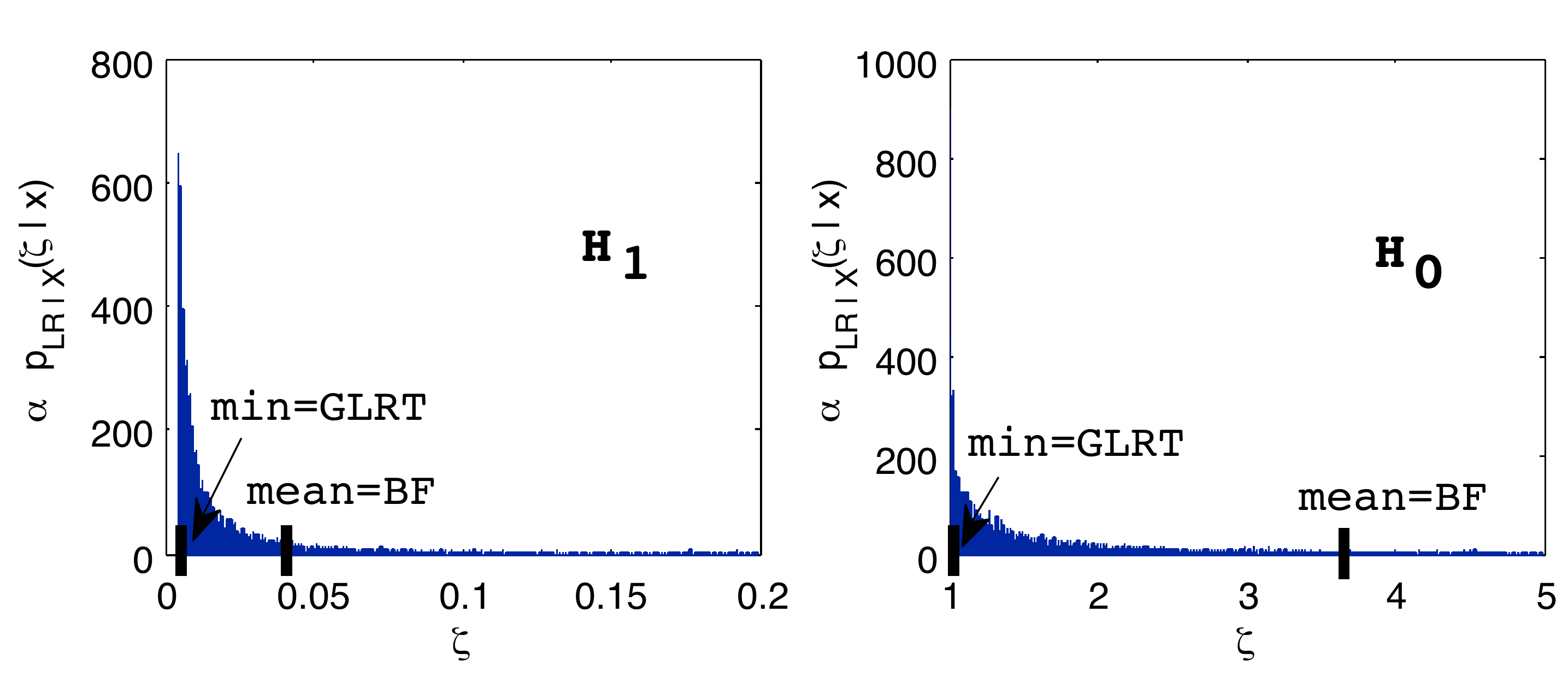}}
\vspace{-0.3cm}
\caption{Histograms of the LR($\bd\eta^{[n]}$) chains, computed from the chains $\bd\eta^{[n]}$ shown in Fig. \ref{fig_posterior}. The (Bayesian) GLRT (\ref{def_GLRT}) and the BF (\ref{def_bf}) are indicated (see Sec. \ref{sec_stats} and \ref{sec_plr_glrt} for the proofs).}
\label{fig_pdf_LR}
\end{figure}

\begin{figure}
\centerline{\includegraphics[width=1\columnwidth]{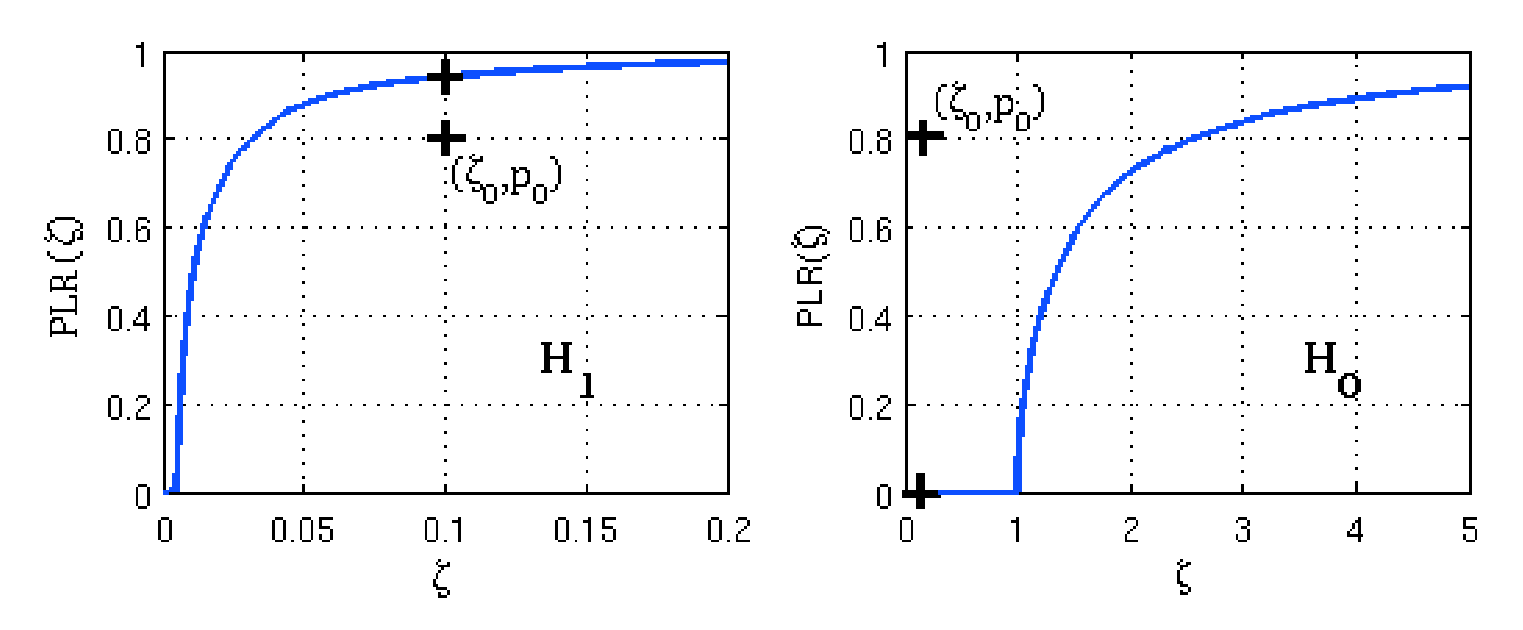}}
\vspace{-0.3cm}
\caption{A posteriori empirical cumulative distributions of LR, displayed from the chains LR($\bd\eta^{[n]}$) shown in Fig. \ref{fig_pdf_LR}.}
\label{fig_PLR}
\end{figure}

Fig. \ref{fig_posterior} shows the histograms of the Markov chains resulting from these two cases. Under the $H_1$ case, the Bayes Factor (\ref{def_bf}) seems to indicate with no ambiguity a detection: BF = 0.04 $< \zeta_0$ for $\zeta_0 = 0.1$. The uncertainty (\ref{def_err}) gives $\hat\sigma$ = 0.34 ("LR = 0.04 $\pm$ 0.34") but for the reasons mentionned in Sec. \ref{sec_stats}, a quantile should be more relevant than a moment to infer the uncertainty on a detector. The measure PLR$(\zeta_0) = 0.94 > 0.8$ confirms the absence of ambiguity of the BF result. Similarly, in the $H_0$ case, the BF test indicates again with no ambiguity that there is no exoplanet: "LR = 3.7 ($\pm$ 86)". This is confirmed by the quantile PLR$(\zeta_0)$ = 0. For a more complete information, the empirical posterior distributions of LR$(\eta^{[n]})$ are presented on Fig. \ref{fig_pdf_LR} and Fig. \ref{fig_PLR}. They also illustrate the properties shown in Sec \ref{sec_stats} and \ref{sec_plr_glrt}. 

Finally, estimation can be performed for the data where a signal has been detected (ie data simulated under $H_1$). The posterior distribution is shown on Fig. \ref{fig_posterior} (left). The signal is estimated by the posterior mean and its uncertainty by the posterior standard deviation: $\bd{\hat \eta} = (6.2 \pm 2.8~ ; 4.6 \pm 2.6).10^{-4}$ for a true $\bd \eta_{H1} = (8 ~; 0.5).10^{-5}$.

\subsection{Comparison with a practical and totally frequentist GLRT}

The proposed procedure is compared to a classical Generalized Likelihood Ratio Test (not the "Bayesian" GLRT (\ref{def_GLRT})). The likelihood used to compute it is the first level likelihood (\ref{likelihood}), except that the covariance matrix is assumed to be proportionnal to identity: $\Sigma = \sigma^2 I_{LM}$. Then,
\begin{align}
\text{GLRT}_2 = \frac{\max_{\bd \mu, \sigma}\{\prod_k p(\bd x_k|\bd \mu,\sigma^2 I_{LM},\bd \eta=\bd 0)\}}{\max_{\bd \mu, \sigma, \bd \eta}\{\prod_k p(\bd x_k|\bd \mu,\sigma^2 I_{LM},\bd \eta)\}} 
\end{align}
The analytical maximization of the likelihood under $H_1$ for $L=2$ generalizes a computation in \cite{smith09} where $L=1$, and leads to:
\begin{align}
\text{GLRT}_2 &= \left(\frac{\widehat \sigma_{H1}^2}{\widehat \sigma_{H0}^2}\right)^{\frac{2KM}{2}}  \label{GLRT} \\
\text{where }~~ \widehat \sigma_{Hi}^2 &= \frac{\sum_k \| \bd x_k - A_k \widehat{\bd \eta}_{Hi} - \widehat{\bd \mu}_{Hi}\|^2}{KLM} \nonumber 
\end{align}
where $\widehat{\bd \eta}_{H0} = \bd 0$ and $(\widehat{\bd \mu}_{H1}^t,\widehat{\bd \eta}_{H1}^t)$ and $\widehat{\bd \mu}_{H0}$ minimize least square criteria obtained from the model (\ref{likelihood}).

The GLRT -contrary to LR($\bd\eta^{[n]}$)- has always a value inferior or equal to 1 because the hypotheses are nested. Here, $\ln(\text{GLRT}_2) = -4350$ for the data simulated under $H_1$ and $\ln(\text{GLRT}_2) = -1300$ under $H_0$. Since it is not numerically possible to realistically simulate a large number of datasets, it is impossible to relate numerically the threshold of the GLRT to its Probability of False Alarm (PFA). The model (\ref{likelihood}) is not identically distributed, so the classical results on the asymptotic distribution of the GLRT neither apply. It is consequently difficult to choose the threshold $\zeta$. 

In any case, the values of the GLRT$_2$ applied to areas closed but distinct from the precedent cases indicate that the GLRT$_2$ discriminates with difficulty $H_0$ and $H_1$.

\section{Illustration of the FBF and PLR optimizations as detectors}
\label{sec_perfs}

The other interesting property presented in Sec. \ref{sec_optim} of the PLR (\ref{def_PLR}) and FBF (\ref{def_FBF}) is now illustrated on an astrophysical context totally similar to the previous one, but the data are now simulated from the statistical model and not the physical one, so that a long run performance analysis can be performed. The data are simulated from the marginalized likelihood presented in \cite{smith09b} for $KLM = 80$. For simplicity, the data under $H_1$ are characterized by a fixed $\bd \eta = \bd \eta_{H1}$.

\begin{figure}[h]
\centerline{\includegraphics[width=0.75\columnwidth]{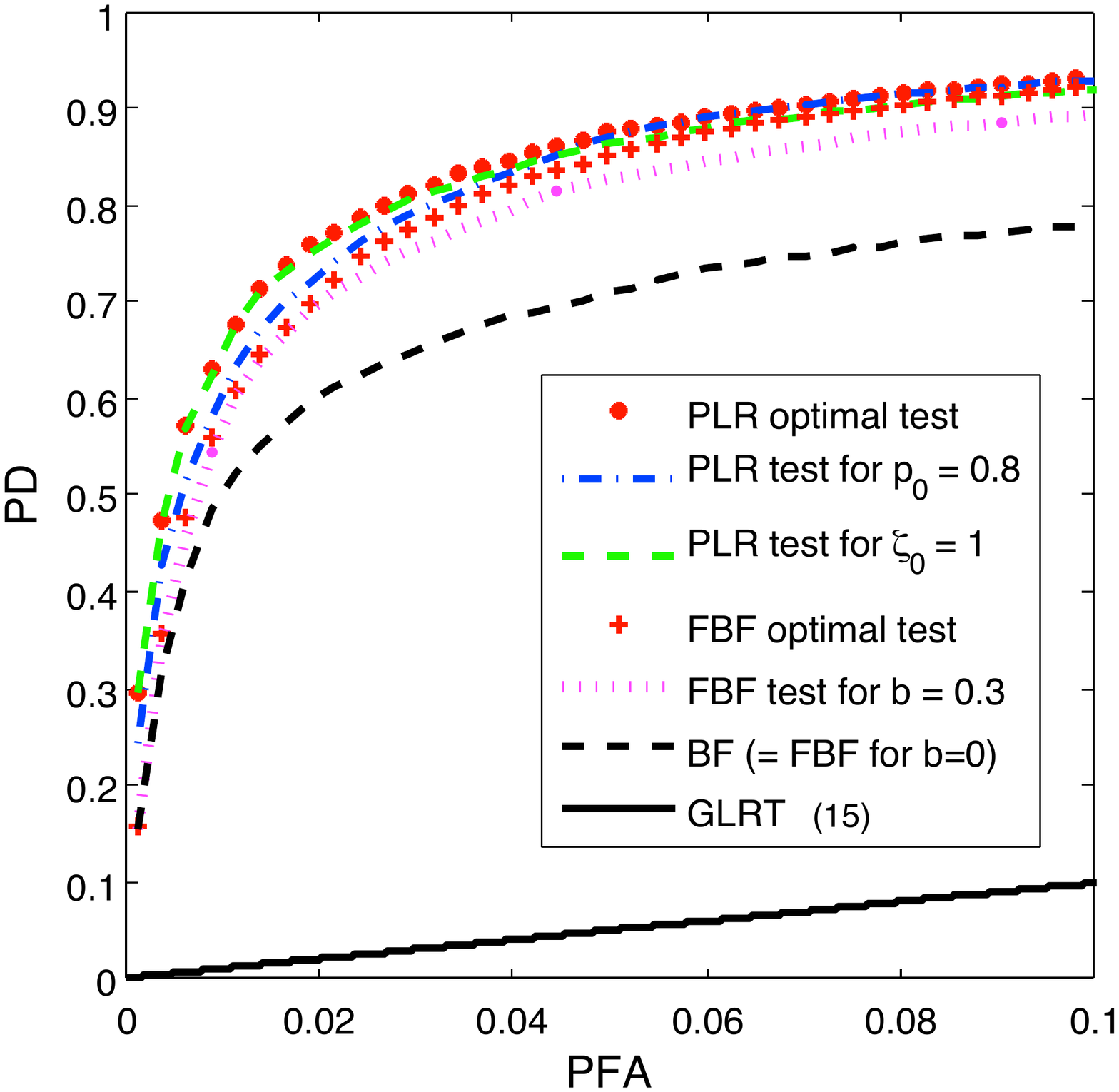}}
\caption{ROC curves of the PLR, the FBF and the GLRT$_2$ (Eq. \ref{GLRT}).}
\label{fig_roc}
\end{figure}

Fig. \ref{fig_roc} illustrates the ROC curves obtained for some intuitive parametrizations ($\zeta_0 = 1$ etc) and the optimal ones. We note that: 
\begin{itemize}
\item The classical Bayes Factor is uniformly less performant than the other FBF and the PLR. For $\text{PFA} = 0.1$, the performances of the PLR overpass the ones of the Bayes Factor by 15\%. 
\item The tests with fixed parametrization have performances very close to the optimal ones. It strenghtens their use.
\item The bad performances of the GLRT (\ref{GLRT}) where $\Sigma$ was wrongly  assumed to be proportionnal to identity are confirmed here: it is equivalent to a heads or tails test.
\end{itemize}

\section{Conclusion}

In this paper, a coherent and practical detection procedure has been proposed. The procedure relies on the fact that for a simple versus composite test using a proper prior the Bayes Factor can be expressed as the posterior mean of the Likelihood Ratio. The statistics involved (FBF and PLR) are computable from the single $\bd \eta^{[n]}\sim\pi^*(\bd\eta|\bd x)$ Markov Chain. It has been realistically applied and compared to a reasonable alternative and proved satisfactory. Finally, two more properties of the PLR -related to the GLRT- have been given, and the PLR and FBF families have been studied as optimizable detectors from a ROC curve. These results have been applied and show that intuitive parametrizations of these tests are close to optimal.

\end{document}